\documentstyle[12pt,aasms4]{article}

\received{June 16 2000}
\accepted{July 26 2000}

\begin{document}

\title{A comparison of central temperatures of the intracluster gas
determined from X-ray and SZ measurements} 

\author{Tong-Jie Zhang}
\affil{Department of Astronomy, Beijing Normal University,
Beijing, 100875; and 
National Astronomical Observatories,
Chinese Academy of Sciences, Beijing 100012, China}

\and

\author{Xiang-Ping Wu}
\affil{Beijing Astronomical Observatory and 
National Astronomical Observatories,
Chinese Academy of Sciences, Beijing 100012, China}

\begin{abstract}
A combination of the X-ray imaging and SZ measurements of clusters
permits an indirect determination 
of the radial temperature profiles of intracluster gas, which requires no
assumption about the dynamical properties and the equation of state 
for clusters. A comparison of such a derived gas temperature with that 
given by the X-ray spectral analysis constitutes an effective probe of 
properties of intracluster gas.
Using the available data of 31 clusters in literature, we have performed
the first comparison of the central gas temperatures provided by the
two methods. The good agreement between these two
independent temperature estimates suggests that  the distribution of
intracluster gas is essentially consistent with isothermality  
characterized by a mean polytropic index of $\gamma=0.9\pm0.1$.
\end{abstract}


\section{Introduction}

While the X-ray spectroscopic measurement is a powerful and unique 
way nowadays to obtain the temperature of hot, diffuse gas contained 
within clusters of galaxies, 
it yields an emission-weighted temperature rather than
the true temperature distribution.  The lack of detailed information
on the temperature profiles in clusters is probably the major source of
uncertainties in the present determinations of the total masses and
baryon fractions of clusters, which further hinders the dynamical
properties of clusters from the cosmological applications such as the
estimates of the cosmological parameters and the test of various
models of structure formation in the universe.
An independent, complementary method 
of measuring the distribution of intracluster gas is thus desirable. 
Silk \& White (1978) were the first to suggest the utilization of 
the nonparametric reconstruction of the radial profiles of density 
[$n_e(r)$] and temperature [$T(r)$] of intracluster gas 
by combining the X-ray and SZ measurements. The basic idea is to 
inverse the observed X-ray and SZ temperature surface brightness profiles of 
clusters which have the functional forms of $n_e^2(r)T^{1/2}(r)$ and 
$n_e(r)T(r)$,  respectively. And a simple combination of the two 
functions gives straightforwardly the gas density $n_e(r)$ and temperature
$T(r)$.  This method requires no assumption about the dynamical
properties of clusters and the equation of state for intracluster
gas, and can therefore be regarded as an ideal and ultimate tool to probe
the gas distribution in clusters under spherical approximation.

Despite its elegant mathematical treatment further developed by 
Yoshikaya \& Suto (1999) based on some theoretical models, 
the pioneering suggestion of Silk \& White (1978)
has not yet been put into practice to date. 
This has been primarily limited by the instrumental sensitivity and 
resolution of detecting the temperature variations of the 
cosmic background radiation (CBR) behind clusters.
So far, the marginal detections of the radial SZ temperature distributions 
have been reported only for a few clusters (see Birkinshaw 1999), 
which can hardly be used for the purpose of reconstructing 
the gas temperature profiles because of the sparse data points 
and the large associated uncertainties.
Nevertheless, we notice that the central measured or estimated 
SZ effects have been available for a great number of clusters 
with the past two decades' efforts, and these central CBR temperature
decrements in the Rayleigh-Jeans limit are usually used for setting
constraints on the cosmological parameters ($H_0$, $\Omega_M$ and
$\Omega_{\Lambda}$) in conjunction with the X-ray imaging and 
spectral measurements. Yet, another application of 
the central SZ data is instead to estimate the central temperature of 
intracluster gas when combined with the X-ray imaging observation, 
although this does not
completely achieve the original goal of Silk \& White (1978). 
It provides an indirect measurement of the X-ray temperature at cluster
centers, which can be  directly compared with the result given by the X-ray 
spectral analysis. Additionally, because both X-ray emission and SZ 
effect result from the gas distribution projected along the
light of sight, a comparison of the central temperatures indicated
by the two methods may also reveal valuable information about the
radial temperature variations. Using the published X-ray and SZ data
in literature, we present for the first time such a comparison
in this paper and discuss its implications for dynamical 
properties of clusters. Throughout the 
paper we assume $H_0=50$ km s$^{-1}$ Mpc$^{-1}$ and $\Omega_0=1$.

\section{The model}

The X-ray imaging observation provides a reliable measurement of the
X-ray azimuthally-averaged surface brightness profile of a cluster
which is usually described by the conventional $\beta$ model
(Cavaliere \& Fusco-Femiano 1976)
\begin{equation}
S_x(r)=S_0\left(1+\frac{r^2}{r_c^2}\right)^{-3\beta+1/2}.
\end{equation}
In the scenario of an optically-thin, thermal bremsstrahlung emission, 
the above form of $S_x(r)$ indicates (Cowie, Henriksen \& Mushotzky 1987)
\begin{equation}
n_e(r)T^{1/4}(r)=n_{e0}T_0^{1/4}
\left(1+\frac{r^2}{r_c^2}\right)^{-3\beta/2},
\end{equation}
where $n_e$ and $T$ are the electron number density and temperature,
respectively. If we assume an equation of state, 
$T(r)=T_0[n_e(r)/n_{e0}]^{\gamma-1}$, we can write the 
electron number density as $n_e(r)=n_{e0}(1+r^2/r_c^2)^{-\delta}$,
in which $\delta=6\beta/(3+\gamma)$. The central electron number
density $n_{e0}$, temperature $T_0$ and the X-ray surface brightness $S_0$ 
are connected by 
\begin{equation}
n_{e0}^2=  \left(\frac{3m_e\hbar c^2}{2^4e^6}\right)
         \left(\frac{3m_ec^2}{2\pi kT_0}\right)^{1/2}
           \frac{4\pi^{1/2}}{\mu_e g}
        \frac{\Gamma(3\beta)}{\Gamma(3\beta-1/2)}
        \frac{S_0(1+z)^4}{r_c},
\end{equation}
where $\mu_e=2/(1+X)$, $X=0.768$ is the primordial hydrogen mass fraction,
$g\approx1.2$ is the average Gaunt factor, $\Gamma$ is the gamma function, 
and $z$ is the cluster redshift.

On the other hand, the CBR temperature decrement/increment at the cluster 
center predicted by the SZ effect is
\begin{eqnarray}
\frac{\Delta T_{sz}}{T_{\rm CBR}}=-4\xi(x)\int_{0}^{+\infty}
           \frac{kT(r)}{m_ec^2}\sigma_T n_e(r) dr,\\
\xi(x)=\frac{x^2e^x}{2(e^x-1)^2}\left(4-x\coth \frac{x}{2}\right),
\end{eqnarray}
where $T_{\rm CBR}$ is the temperature of the present CBR, and  
$x=h\nu/kT_{\rm CBR}$ is the dimensionless frequency. 
A straightforward computation using eqs.(2)-(5) as well as 
the polytropic equation of state gives
\begin{eqnarray}
\left(\frac{kT_0}{m_ec^2}\right)^{3/2}= &
        \left(\frac{\Delta T_{sz}}{T_{\rm CBR}}\right)^2
        \left(\frac{\Gamma(3\beta^{\prime}/2)}
             {\Gamma(3\beta^{\prime}/2-1/2)}\right)^2 
	   \frac{\Gamma(3\beta-1/2)}{\Gamma(3\beta)} \nonumber\\
	& \;\cdot\; \frac{\sqrt{2} \alpha \mu_e g m_e c^3 }
	     {8\pi^2\sqrt{3}(3\beta-3/2)\xi^2\sigma_T(1+z)^4S_0r_c}
\end{eqnarray}
in which $\beta^{\prime}=3\beta\gamma/(3+\gamma)$, 
and $\alpha$ is the fine structure
constant. The above expression is often used in the determination of
the Hubble constant for an isothermal gas distribution 
if we write $r_c=d_A(z)\theta_c$ where $d_A$ is the
angular diameter distance to the cluster at $z$. On the other hand, 
given the Hubble constant and the polytropic index $\gamma$,  
all the parameters at the right-hand of eq.(6) 
are measurable from the X-ray imaging and SZ measurements, which 
will instead allow us to derive the central temperature of 
the X-ray emitting gas.

\section{Application to X-ray clusters}

Our cluster sample consists primarily of 21 nearby clusters  
studied extensively by Mason \& Myers (2000) plus 10 high-redshift
clusters with recent, reliable SZ data in the literature.  
Inclusion of a cluster is also subject to whether or not 
the good spectral temperature data are available
in order to facilitate our comparison between different methods.
We take the best-fit values of $\beta$, $r_c$ and $S_0$ (or $L_x$)
as well as the emission-weighted temperature data 
directly from the literature. For the latter, we adopt
the temperature given by a single-phase (s.p.) model and 
the value by excluding the cooling flows (c.l.) (White 2000),
respectively. Since the X-ray spectrum is always
observed over a finite bandpass, we assume an optically-thin, 
thermal bremsstrahlung emission model with the metal abundances of 
$30\%$ solar to convert the observed $S_0$ (or $L_x$) over the 
corresponding energy band into the bolometric value.   
We use the central SZ decrement/increment  of $\Delta T_{sz}$ 
after the correction of the finite beam in each observation.
All the input data of the X-ray and SZ measurements are summarized in
Table 1.

For an isothermal gas distribution, the jointly determined central 
temperature, $T_{est}$, from the X-ray imaging and SZ measurements
(eq.[6]) can be directly compared with the X-ray spectral result, $T_{spec}$.
We display in Fig.1 our derived $T_{est}$ versus the observed 
$T_{spec}$ (s.p. model) 
for the 31 clusters in Table 1. A glimpse of Fig.1 reveals
that the two sets of data are roughly consistent with each other.
The best-fit $T_{spec}$-$T_{est}$ relation reads
\begin{eqnarray}
T_{est}=10^{-0.22\pm0.18}T_{spec}^{1.22\pm0.20}, & 
                           {\rm s.p.\; model};\\
T_{set}=10^{-0.36\pm0.26}T_{spec}^{1.29\pm0.29}, & 
                           {\rm c.l.\; corrected},
\end{eqnarray}
in which the error bars are the $90\%$ confidence limits which have 
taken the measurement uncertainties into account.
We have also tested the dependence of the best-fit 
$T_{spec}$-$T_{est}$ relation on the Hubble constant by adopting a larger
value of $H_0=75$ km s$^{-1}$ Mpc$^{-1}$. This only leads to a minor 
modification to the above results. For instance, the best-fit relation for 
the s.p. model now reads $T_{est}=10^{-0.34\pm0.11}T_{spec}^{1.22\pm0.12}$.

\placefigure{fig1}

If the intracluster gas deviates from isothermality, 
our comparison of  $T_{est}$ and $T_{spec}$ 
becomes to be a little complex because we need to recover the central
temperature from the emission-weighted spectral fit. On the
other hand, we can place a useful constraint on the polytropic index 
$\gamma$ if we demand that our derived central temperature from eq.(6) 
identify the spectral value after a proper correction to 
the emission-weighted temperature $\overline{T}$ predicted by 
\begin{equation}
\frac{dE}{d\nu}\propto \overline{T}^{-1/2} g_{ff}(\overline{T},\nu) 
               \exp(-h\nu/k\overline{T})
               \int n_e^2 dV,
\end{equation}
where $g_{ff}$ is the Gaunt factor of the free-free emission, $\nu$ is
the X-ray observing frequency, and the integration is performed over the whole
cluster volume if we neglect the finite detection aperture.
This compares with the theoretically expected spectrum for 
the hot intracluster gas
\begin{equation}
\frac{dE}{d\nu}\propto \int n_e^2 T^{-1/2} g_{ff}(T,\nu)
                       \exp(-h\nu/k{T}) dV.
\end{equation}
For each cluster with the reported $\overline{T}=T_{spec}$, 
we use the Monte Carlo simulations over the typical energy band 
$0.5$ - $10$ keV to obtain the central temperature $T_0$ and the
corresponding $\gamma$ with the restriction of eq.(6). The
resultant $\gamma$ value for each cluster is listed in Table 1, 
and a combined analysis of the results among the 31 clusters yields
\begin{eqnarray}
\langle \gamma\rangle =0.91\pm0.14, &  {\rm s.p. \; model};\\
\langle \gamma\rangle =0.93\pm0.14, &  {\rm c.l.\; corrected},
\end{eqnarray}
which indicates that clusters are essentially consistent with isothermality
at $90\%$ significance level.

\section{Discussion and conclusions}

Our joint analysis of the indirect and direct temperature measurements 
from the X-ray and SZ observations provides a useful clue to resolving
the temperature profile discrepancy raised in recent years 
particularly by Markevitch et al. (1998), Irwin, Bregman \& Evrard (1999), 
White (2000) and Irwin \& Bregman (2000). Markevitch et al. (1998) claimed
a significant temperature decline with increasing radius represented
by a polytropic index of 1.2-1.3, while the other investigators have
essentially detected an isothermal temperature profile.
Our present study gives an independent yet convincing evidence 
for the assumption of isothermality. The advantage of our algorithm
over the previous analysis is that we have obtained the central temperatures
of clusters without utilizing the X-ray spectral data, while a combination
of both has allowed us to set a valuable constraint on the radial temperature
profiles of the hot X-ray emitting gas.

A joint determination of the central temperature of intracluster gas 
from the X-ray imaging and SZ measurements can be used at present 
as an independent and complementary method to examine the reliability and
accuracy of the temperature
measurement provided by the conventional X-ray spectral analysis,
although our ultimate goal is to obtain the detailed radial profiles of 
electron number density and temperature when the SZ surface brightness
can be precisely mapped (Silk \& White, 1978; Yoshikawa \& Suto 1999).
Yet, we should point out that it may not be easy to identify
the temperature structure of the hot gas from such a combined analysis,
in the sense that the dominant component in both $S_x$ and $\Delta T_{sz}$
is characterized by the sharp decrease of 
the electron number density with outward radius, while it is unlikely that
the temperature profiles can demonstrate a similar variation.
An accurate measurement of $S_x$ and $\Delta T_{sz}$ at outermost radii 
will thus be needed if one wants to apply the present technique to 
the reconstruction of temperature profiles of intracluster gas.
Alternatively, the present technique requires that the X-ray surface 
brightness is well represented by the $\beta$ model. Presence of cooling
flows in the central regions of many X-ray clusters may yield an 
underestimate of the $\beta$ parameter and the core radius $r_c$ and
an overestimate of the central surface brightness $S_0$,
which will in turn affect our estimate of $T_{est}$ through eq.(6).
Indeed, an examination of Fig.1 reveals that all the massive 
cooling flow clusters with $\dot{M}>600$ $M_{\odot}$ yr$^{-1}$ 
(A478, A1835, A2204 and  Zw3146) demonstrate rather large scatters around
the $T_{est}=T_{spec}$ relation. Therefore, the present method is most 
applicable to morphologically simple clusters without cooling flows.

\acknowledgments
This work was supported by 
the National Science Foundation of China, under Grant 19725311.

\clearpage

\figcaption{The derived central temperatures are plotted against 
the spectral results (single-phase model) for a sample of 31 clusters. 
Solid line shows the best-fit relation, and dotted
line indicates $T_{est}=T_{spec}$. The high-redshift ($z>0.1$) and 
low-redshift ($z<0.1$)
clusters are represented by the filled squares and triangles,
respectively. The marked clusters are the massive cooling flow ones with 
$\dot{M}>600$ $M_{\odot}$ yr$^{-1}$ (White 2000). 
\label{fig1}}

\end{document}